\documentclass[amt, manuscript]{copernicus}
\usepackage{siunitx}

\begin{document}
\nolinenumbers
\title{Mimicking non-ideal instrument behavior for hologram processing using neural style translation}

\Author[1]{John S.}{Schreck}
\Author[2]{Matthew}{Hayman}
\Author[1]{Gabrielle}{Gantos}
\Author[3]{Aaron}{Bansemer}
\Author[1]{David John}{Gagne}

\affil[1]{National Center for Atmospheric Research (NCAR), Computational and Information Systems Laboratory, Boulder, CO, USA}
\affil[2]{National Center for Atmospheric Research (NCAR), Earth Observing Laboratory, Boulder, CO, USA}
\affil[3]{National Center for Atmospheric Research (NCAR), Mesoscale and Microscale Meteorology Laboratory, Boulder, CO, USA}

\correspondence{schreck@ucar.edu; mhayman@ucar.edu}

\runningtitle{Mimicking non-ideal instrument behavior for hologram processing using neural style translation}

\runningauthor{Schreck et al.}

\received{}
\pubdiscuss{} 
\revised{}
\accepted{}
\published{}


\firstpage{1}

\maketitle

\begin{abstract}

Holographic cloud probes provide unprecedented information on cloud particle density, size and position. Each laser shot captures particles within a large volume, where images can be computationally refocused to determine particle size and shape. However, processing these holograms, either with standard methods or with machine learning (ML) models, requires considerable computational resources, time and occasional human intervention. Current ML models are trained on simulated holograms obtained from the physical model of the probe since real holograms have no absolute truth labels. Any attempt to use another processing method to produce labels would be subject to errors that the ML model would subsequently inherit. Trained models perform well on real holograms only when image corruption is performed on the simulated images during training, thereby mimicking non-ideal conditions in the actual probe (Schreck et. al, 2022). Optimizing image corruption requires a cumbersome manual labeling effort. 

Here we demonstrate the feasibility of applying the neural style translation approach (Gatys et. al, 2016) to the simulated holograms. With a standard, pre-trained convolutional neural network (VGG-19), the simulated holograms are “stylized” to resemble the real ones obtained from the probe, while at the same time preserving the simulated image “content” (e.g. the particle locations and sizes). Several image similarity metrics concur that the stylized images are more like real holograms than the synthetic ones. With an ML model trained to predict particle locations and shapes on the stylized data sets, we observed comparable performance on both simulated and real holograms, obviating the need to perform manual labeling. The approach described here is not specific to hologram images and could be applied in other domains as a means for capturing noise and imperfections in observational instruments to make simulated data more like real world observations.




\end{abstract}

\introduction
Holographic imaging enables 3D reconstruction of particle positions and sizes within a large sample volume.  Through this technique, ensembles of thousands of particles can be captured and characterized in a single instantaneous volume.  This technique has been successfully applied to characterizing the microphysical properties of clouds with airborne in situ instruments such as the HOLODEC probe \citep{fugal2004airborne,spuler2011design} which is operated by NCAR and has been successfully deployed on several field campaigns.

The HOLODEC instrument captures inline holograms of a 15 $cm^3$ sample volume.  A laser operating at 355 nm transmits an expanded beam between two probe arms.  After the light passes through the sample volume, the light is imaged onto a CCD with an effective pixel resolution of 3 $\mu m$.  The laser fires at a 3 Hz repetition rate, allowing for point-like captures of particle size distributions as the aircraft passes through a cloud with spatial separations of those points dictated by aircraft speed.

While holographic imaging captures a significant amount of information in a single image, processing these holograms tends to be computationally intensive, often requiring human intervention to tune the processing approach.  As a result, there has been substantial research into the use of machine learning to accelerate or improve processing of holographic particle data \citep{Shimobaba2019,Shao2020,zhang2022holographic}.  However, many of these efforts are focused on images with relatively small depth components which is distinct from HOLODEC requirements.  Another key distinction is in the idyllic nature of the holograms that are processed.  Processing solutions that work well on controlled laboratory or simulated holograms tend to perform suboptimally on operationally captured data from HOLODEC.  One cannot expect the same consistent clarity and quality from an instrument on the outside of an aircraft, where a variety of operating and environmental conditions can cause the captured images to contain a significant amount of imperfections.  The challenge of processing HOLODEC data necessitates tuning filters to minimize both false positives and missing true particles in holograms with non-negligible imperfections.

These imperfections are difficult to model in simulation because they often depend on combinations of component imperfections within the instrument.  For example, small amounts of reflections from transmission optics can create modulation patterns and focus spots in reconstruction, laser modes deviate in amplitude and phase from a true Gaussian, vignetting can cause non-uniform field response and depend on beam pointing that varies with vibrations and temperature.  Ultimately, developing processing techniques that can account for these and other (sometimes unknown) effects is essential to delivering high quality scientific observations.

In a previous work, we outlined a processing pipeline concept to accelerate and improve processing of HOLODEC images through a combination of hardware acceleration and machine learning \citep{schreck2022}.  As is common in atmospheric sensing, actual hologram data lacked truth labels.  In order to achieve an improvement in processing capability over the existing solution, we developed an approach for corrupting simulated holograms in order to emulate the imperfections of actual HOLODEC data.  A machine learning solution to process actual HOLODEC data was trained using simulated holograms (where labels are are known to near absolute accuracy) by adding random noise, applying blurring and skewing, and adjusting the image contrast.  While the simulated training data did not visually appear similar to actual HOLODEC data, these image corruptions were sufficient to train a model that still out performed the particle recognition approaches used in the standard processing solution \citep{schreck2022}.  However, to succeed, each of these transforms needed to be tuned in hyperparameter optimization, which required manual labeling of thousands of reconstructed holograms. This is problematic because manual labeling requires considerable person-hours from well trained personnel, and (as we discovered in the process) human determined labels have some non-zero but unquantified error. What is more, during the optimization, we simply guessed which types of corruptions to perform on the images. Finally, there may be many sensor situations where these simple corruption operations do not suffice in developing a higher performance processing solution. In order to generalize to the broader domain of sensor signal processing, a solution should be able to address the specialized behavior of the sensor type and this mandates a more general ability to represent the difficult to model, non-ideal behaviors of the instrument.

Here we introduce a new approach to bridging the gap between optimizing signal processing/machine learning solutions and the realities of operational sensor data. We show that the original neural style transfer method introduced by \citep{gatys2016}, which utilizes a convolutional neural network (CNN) to perform image augmentation, can be used to produce realistic-looking, labeled hologram examples without the need for manually labeling as a means for tuning noise parameters. These stylized synethetic training data result in processing performance similar to the previous method, but without the requirement of human labeling and therefore represent a significant step toward producing an operation solution for processing HOLODEC data.  

\section{Methodology}

\subsection{Neural Style Transfer}

\begin{figure}[t!]
    \centering
    \includegraphics[width=\columnwidth]{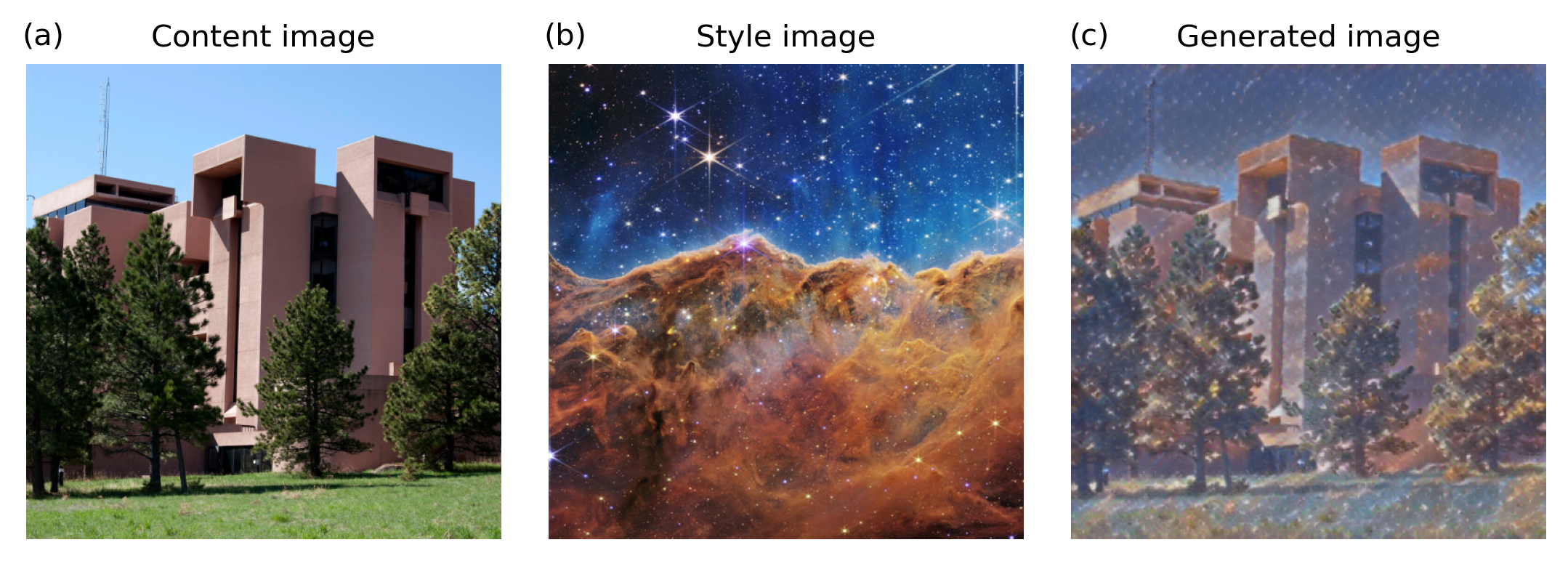}
    \caption{(a) NCAR's Mesa Lab taken to be the content image \citep{mesa_lab}. (b) An image obtained from the James Webb telescope showing star formation is the style image \citep{james_webb}. (c) The synthesized image is the product of the neural style transfer method that leverages the content and style images.}
    \label{fig:mesa_lab}
\end{figure}

The original style transfer method leverages CNN representations of a ``content'' image and a ``style'' image to produce a third image where the style features are automatically synthesized with content features. For example, Figure~\ref{fig:mesa_lab} illustrates how the method is used to create an image of NCAR's Mesa Lab shown in (a), stylized using a recent image from the James Webb telescope shown in (b) to create a synthesized image shown in (c). The synthesized image still clearly still shows the Mesa Lab, however, overall the image colors, contrast, and line smoothnesses have been transformed relative to the original to also resemble the style image. 

The synthetic holograms come with labels and are free of noise, so they are identified as possessing the ``content'' we wish to preserve --- that is the information about the particles within. The HOLODEC examples all contain varying degrees of noise/imperfections that we wish to effectively impart onto the synthetic examples, hence they are identified as the ``style'' examples.  In this way, synthetic images are transformed to create a realistic training dataset with known labels.

\begin{figure}[t!]
    \centering
    \includegraphics[width=6 in]{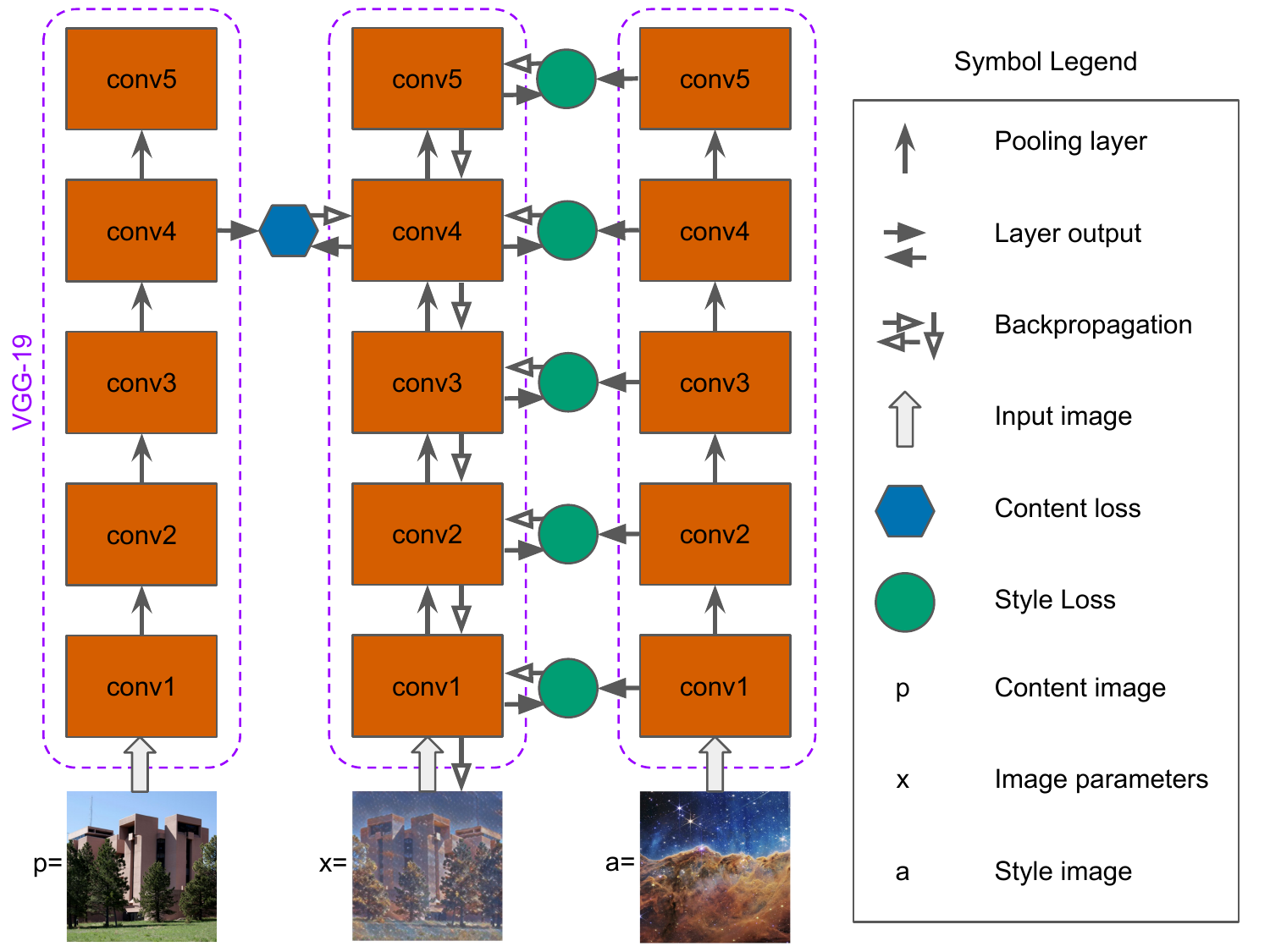}
    \caption{Schematic illustration of the neural style transfer approach by \cite{gatys2016} using the VGG-19 neural model, where the model's 18 convolution layers are stacked into five blocks (orange squares). In this example, the output from the first convolution layer in each of the five blocks were used to compute style and content losses. The blocks represent the (same) model for inputs $p$, $x$, and $a$.}
    \label{fig:style_model}
\end{figure}

Figure~\ref{fig:style_model} schematically illustrates the style transfer method using the VGG-19 network as the chosen CNN \citep{vgg}. The VGG model contains 19 layers and was originally trained to perform object recognition and localisation on the ImageNet dataset \citep{imagenet}. The model's 18 convolution layers utilize the ReLU activation and are sequentially connected together into five blocks separated by 2D average Pooling layers, as is illustrated schematically in Figure~\ref{fig:style_model}. The first two blocks contain two convolution layers, while the remaining three contain four. The position of a convolution layer in VGG is characterized by integer $l$. All convolution layers in VGG-19 use a kernel size of $3 x 3$ and stride and padding sizes both set to unity. Three fully-connected layers complete VGG-19. Thus, each column in Figure~\ref{fig:style_model} is a single (the same) VGG-19 model (with the fully-connected layers discarded). Currently, a single color channel is used for the hologram images, so the very first convolution layer was replaced with one having the color channel equal to one. 

The style transfer approach begins by sampling random values for $x$, and freezing the weights of the all of the layers in the CNN. For the input content image $p$ and generated image $x$, the output from convolution layer $l$ are the feature matrix representations $P^l$ and $F^l$, where $P_{ij}^l$ and $F_{ij}^l$ are the responses to the activation of the $ith$ filter at position $j$ in convolution layer at position $l$. A central question with style-transfer is the choice of loss function, and which layer outputs should be used to compute the loss. The choice will determine to what extent we need the full VGG-19 model, and will greatly affect run-time performance. We follow the original approach introduced by \cite{gatys2016}, but other effective loss functions have been developed \citep{li2016_wand, li2017, wilmot2017}. A new image that matches the responses of the content image $p$ may be found by updating the values of $x$ via gradient descent using a squared-loss of the two feature representations 
\begin{equation}
    \mathcal{L}_{content}(p, x, l) = \frac{1}{2} \sum_{i,j} (F_{ij}^l - P_{ij}^l)^2
\end{equation}
for convolution layer at position $l$, as is illustrated in Figure~\ref{fig:style_model} by a blue hexagon. The ``style'' representation of an input image $x$ is defined using a feature space for capturing texture information. This is taken to be the correlation between the outputs across the spatial domain for layer $l$, so that the expected correlation for style feature maps $i$ and $j$ are given by the Gram matrix 
\begin{equation}
    G_{ij}^l = \sum_{k} G_{ik}^l G_{jk}^l.
\end{equation}
Now the image $x$ may be updated via gradient descent so that it's style features $G^l$ match the style features $A^l$ of image a at layer $l$. The contribution coming from layer $l$ is taken to be
\begin{equation}
    E_l = \frac{1}{4 N_l^2 M_l^2} \sum_{i,j} (G_{ij}^l - A_{ij}^l)^2
\end{equation}
and the total style loss, illustrated by green circles in Figure~\ref{fig:style_model}, is
\begin{equation}
\label{style_eq}
    \mathcal{L}_{style}(a, x) = \sum_{l} w_l E_l
\end{equation}
where $w_l$ represents a weight for each contribution coming from layer $l$. Although not used in \cite{gatys2016}, the predicted images may sometimes contain higher-frequency noise which we would prefer to smooth out. This can be accomplished via a total variation term given by
\begin{equation}
    \mathcal{L}_{TV}(x) = \sum_{i,j} \left| x_{i,j}-x_{i+1,j}\right| + \left| x_{i,j}-x_{i,j+1} \right|
\end{equation}
for pixels $i$ and $j$. For content image $p$, style image $a$, and generated image $x$, the total style transfer loss is then
\begin{equation}
\label{style_loss}
    \mathcal{L}_{total}(p,a,x) = \alpha \mathcal{L}_{content}(p, x) + \beta \mathcal{L}_{style}(a, x) + \gamma \mathcal{L}_{TV}(x)
\end{equation}
where the parameters $\alpha$, $\beta$, and $\gamma$ are constant weights, and only the output from a single layer is used to compute the content loss (see Figure~\ref{fig:style_model}). In general, if more than one layer output is selected for computing the content loss, the terms would be summed in Equation~\ref{style_loss}. We set $\alpha$ = $\gamma = 1$ and focus on the relationship between $\beta$ and the predicted images below. 

In addition to $\beta$, the choice of which layer(s) to use in content and style representations are left as hyperparameters. For the content feature layer choice, the output from convolution layers farther from the input image should capture global features relative to those blocks nearer to the input image, which should capture more details of the input. Similar arguments apply to the style features, with the more distant convolution layers capturing more global styles and vice versa. \cite{gatys2016} selected the output from the second convolution layer from the fourth block to compute the content loss (e.g. the 12th convolution layer in VGG-19), and the output from the first convolution layer in all five blocks to compute the style loss components (convolution layers at positions 1, 3, 5, 9 and 13). These choices result in a large CNN that greatly affects training performance when used with our relatively large holograms. We found that a much smaller model was just as capable as that used by \cite{gatys2016} and was thus  much faster to use. Specifically, we selected the fourth convolution layer outputs in VGG-19 for the content contributions and the first five convolution layer outputs in VGG-19 for the style contributions (with all weights equal to unity in Eq.~\ref{style_eq}), therefore only requiring layers in three of the five blocks in VGG-19. There may be a more optimal, faster model that achieves similar results, but we did not explore other combinations any further. 

Figure~\ref{fig:style_model} shows that content image $p$ and style image $x$ are passed through the CNN and the outputs from selected feature layers are stored in each case. Initially white noise image $x$ is then passed through the CNN and the outputs from the content and feature layers are used to compute the total loss according to Equation~\ref{style_loss}. The image pixels in $x$ can then be updated via back-propagation and with successive iterations, $x$ can learn to match the $p$'s content and $a$'s style features. 

\subsection{Metrics for comparing images}

The choice of $\alpha$ and $\beta$ will determine how similar the generated image is to the content and style images. There are numerous metrics available for estimating the similarity between two images. For images $x_0$ and $x_1$, we select the structural similarity index measure (SSIM) \citep{ssim} (Equation~\ref{eq:ssim}), a standard pixel-based metric that extracts information on structure, contrast, and luminance of an image that serves as the basis for comparison of two images. Additionally, a similarity metric based on the AlexNet convolutional neural network model \citep{alexnet}, referred to as $d_\text{Alex} \equiv d_\text{Alex}(x_0, x_1)$, is used to capture the distance between two images, $x_0$ and $x_1$, in deep feature space \citep{alexnet_metric}. The SSIM ranges from -1 to 1, with larger values indicating more similarity between images. The AlexNet metric ranges from 0 to 1 with smaller value indicating greater similarity; we use one minus the value so that a higher value indicates greater similarity in both metrics. For images $x_0$ and $x_1$, both metrics are computed by averaging over local differences between ``patches'' of size $11 \times 11$ for SSIM, and size $3 \times 3$ (the filter size used) for the AlexNet metric.

\subsection{Image segmentation with a U-Net}

\begin{figure}[t!]
    \centering
    \includegraphics[width=\columnwidth]{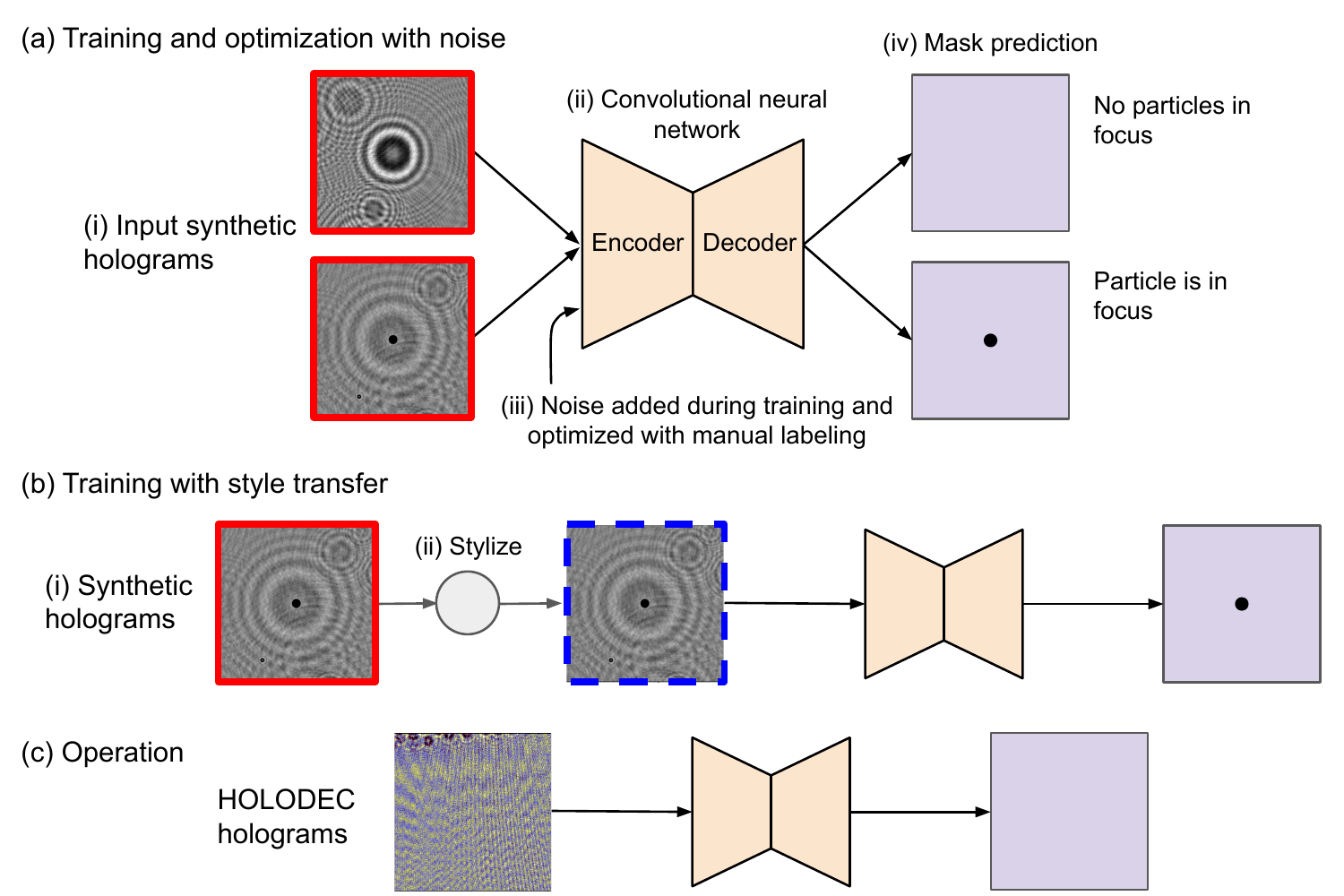}
    \caption{(a) U-Net segmentation model. (i) Example synthetic holograms (solid red lines) are passed into the encoder-head of the U-Net shown in (ii). (iii) During training, noise may optionally be added to the holograms. (iv) Predicted masks for the input examples, with a dark circle enclosing the in-focus particle. (b) (i) Synthetic images are (ii) stylized using HOLODEC holograms (dashed blue lines), which are then (iii) used to train a U-Net model. (c) In operation, HOLODEC holograms are passed through the U-Net trained on style-transferred images to obtain a mask prediction. In the example no particles were identified.}
    \label{fig:unet}
\end{figure}

Figure~\ref{fig:unet} shows a neural segmentation model, for example a ``U-Net'' \citep{ronneberger2015u, segmentation17}, that is used in the current study to process holograms and has been used in other studies \citep{Shao2020}. The model contains two CNN ``heads'' as illustrated in the figure. An encoder head takes in the input image and compresses it into a smaller dimensional representation of a fixed size. The decoder head up-samples the latent representation of the image and outputs a prediction the same size as the input image. Here, we use a sigmoid activation function on the decoder's last layer that rescales the predicted outputs for all the pixels to lie within (0, 1). In other words, the model is trained to predict a probability that an in-focus particle occupies a pixel. If the probability exceeds a certain value, often 0.5 for binary tasks, the pixel is labeled 1 and 0 otherwise.  The convolutional layers in the U-Net encoder head may also leverage pre-trained weights from the layers in other convolutional models, such as ResNet \citep{resnet16} trained on the ImageNet dataset, which frequently helps to speed training and also boost overall performance. 

\subsection{Datasets}
\label{datasets}

There are essentially four datasets we use in this work.  The first two originate from full size images captured by HOLODEC or simulations of HOLODEC.  These initial images consist of 4872-by-3248 pixel images which are processed and segmented to create images that can be readily processed by the U-Net.  These images are also sampled to ensure a balance between in-focus particles and out-of-focus particles in the U-Net training process.  This training data from the two sources are then used by style transfer to generate a realistic generated dataset (with content from the synthetic images and style from the HOLODEC images) which is used to train the U-Net. Here, style transfer is performed using reconstructed synthetic and HOLODEC images at different positions along $z$ (see below). Finally a set of manually labeled HOLODEC holograms are used to evaluate the performance of the U-Net, and therefore the effectiveness of the style transfer solution.

\subsubsection{HOLODEC and synthetic data}
A physical model of the instrument was used to generate 120 synthetic holograms which had optical settings identical to the HOLODEC instrument \citep{fugal2009practical}. One hundred holograms were randomly selected for training, ten for validation, and the remaining ten for testing. Each hologram contained 500 spherically shaped particles randomly positioned along the $x$ and $y$ directions, positioned between minimum and maximum values of 14.072 and 158.928 millimeters along the z-direction, respectively. The diameter sizes were sampled with a gamma distribution.

The HOLODEC data were selected from a set of several hundred holograms obtained from the RF07 subset from the Cloud Systems Evolution in the Trades (CSET) project that was originally obtained from June 1 to Aug 15, 2015 \citep{albrecht2019cloud}. They primarily contain liquid water particles of varying numbers and shape, were largely free of ice but contain other unidentified ``artifacts'' and image perturbations not captured by the physical model of the instrument. Overall, the known particle numbers range from none up to hundreds, with the average particle diameter less than 50 $\textmu m$. 

In accordance with the approach described in \citep{schreck2022}, each hologram is initially processed by reconstructing 1,000 planes along the z-axis (axis of laser propagation).  This produces 1,000, 12 megapixel size images that must scanned for in-focus particles.  These images are too large to be processed efficiently with a CNN. Therefore, following our previous approach \citep{schreck2022}, each hologram reconstruction is broken up into 38-by-28 tiles each of 512 by 512 pixels.  These tiles have an overlap stride of 128 pixels leaving 828 total tiles for each full image. This means that any one pixel will show up in more than 1 tile. 

For training using synthetic holograms, each was selected and then refocused along z for all of the 500 particles, from which a random grid tile containing the in-focus particle was selected and saved, resulting in 50,000 images (for details on refocusing, see \cite{schreck2022}). Additionally, 25,000 tiles were selected where a particle was just out of focus by no more than 120 micrometers. Finally, 25,000 tiles were randomly selected which did not contain in-focus particles. A comparable procedure was repeated to produce validation and test sets of tile images where half the images contained in-focus particles. 

For the RF07 subset used here, a hologram was drawn at random from several hundred, excluding 20 that were set aside previously for manual examination (and described below). This raw data can be accessed at https://doi.org/10.5281/zenodo.6347222. As the true particle positions are not known (and they are not needed as they will be used only for style transfer), the hologram was refocused at a random position between 14 and 160 millimeters along $z$, from which 10 tiles were randomly selected and saved from the refocused hologram. This process was repeated until about 100,000 image tiles were selected for training, 10,000 for validation, and 10,000 for testing the mask prediction performance. Most of the tiles in each split did not have particles directly in-focus, but occasionally were pretty close. 

\subsubsection{Generated data sets with style transfer}

The labeled synthetic training splits are used as the content images while RF07 holograms, which come without labels, are the style images.  Training, validation, and testing splits for 20 different values of $\beta$ sampled evenly from approximately $10^{-5}$ to $10^{15}$, with $\alpha = 1$ and $\gamma = 1$ were created to find the optimal value of $\beta$.  Once obtained, a data set containing training, validation, and testing splits was created with $\beta \equiv \beta_0 = 10^9$, which we identify below as an optimal choice, and $\alpha$ and $\gamma$ both equal to one.  This is created to compare against the model trained on synthetic training split only and the model described in \cite{schreck2022} where optimized noise (based on hand labeled data) was added during training.

For each selection of $\beta$ including $\beta_0$, a randomly selected content (synthetic) image tile was selected and paired off with a randomly selected style (HOLODEC) image tile. Then, both content and style images were used to generate a training sample with the style-transfer method, with the mask labels for the generated example are inherited from the content image. For the $\beta_0$ data set, 100,000, 20,000, and 20,000 style-generated images were created for training, validation, and testing the ability of U-Net model's to reproduce the mask labels. For the other 20 values of $\beta$, a total of 20,000 training, 2,000 validation, and 2,000 testing samples were produced in each case for the same purpose. In the latter data sets, we also used the same random seed so that all the pairs selected in each split for the different $\beta$ choices (excluding $\beta_0$) were the same.

\subsubsection{Manually labeled HOLODEC data set}

The 20 manually labeled HOLODEC examples resulted in 2,356 total images of size 512 pixels by 512 pixels, where prospective in-focus particles were positioned at the image center. The examples were manually labeled as containing an in-focus particle or not, along with an average reviewer confidence score ranging from zero to five (higher means more confident). The binary output of the U-Net is simply taken to be unity if any predicted value is larger than 0.5 and 0 otherwise. Many of the examples were subjective, and the total of number of particles we could identify was limited by previous modeling attempts, hence there could be more particles in these holograms still undiscovered.  

The total number of labeled examples was split into validation and testing  data sets. The validation set contained 1,204 total images with 367 containing at least one in-focus particle. The testing set contained 1,154 images, where 874 contained at least one in-focus particle. The validation set was used to help guide the optimization of noise in our previous study that we compare with here \citep{schreck2022}, while the test set represents the hold-out set of manually labeled examples. 

\section{Results}

\subsection{Style transfer with holograms}

\begin{figure}[t!]
    \centering
    \includegraphics[width=6 in]{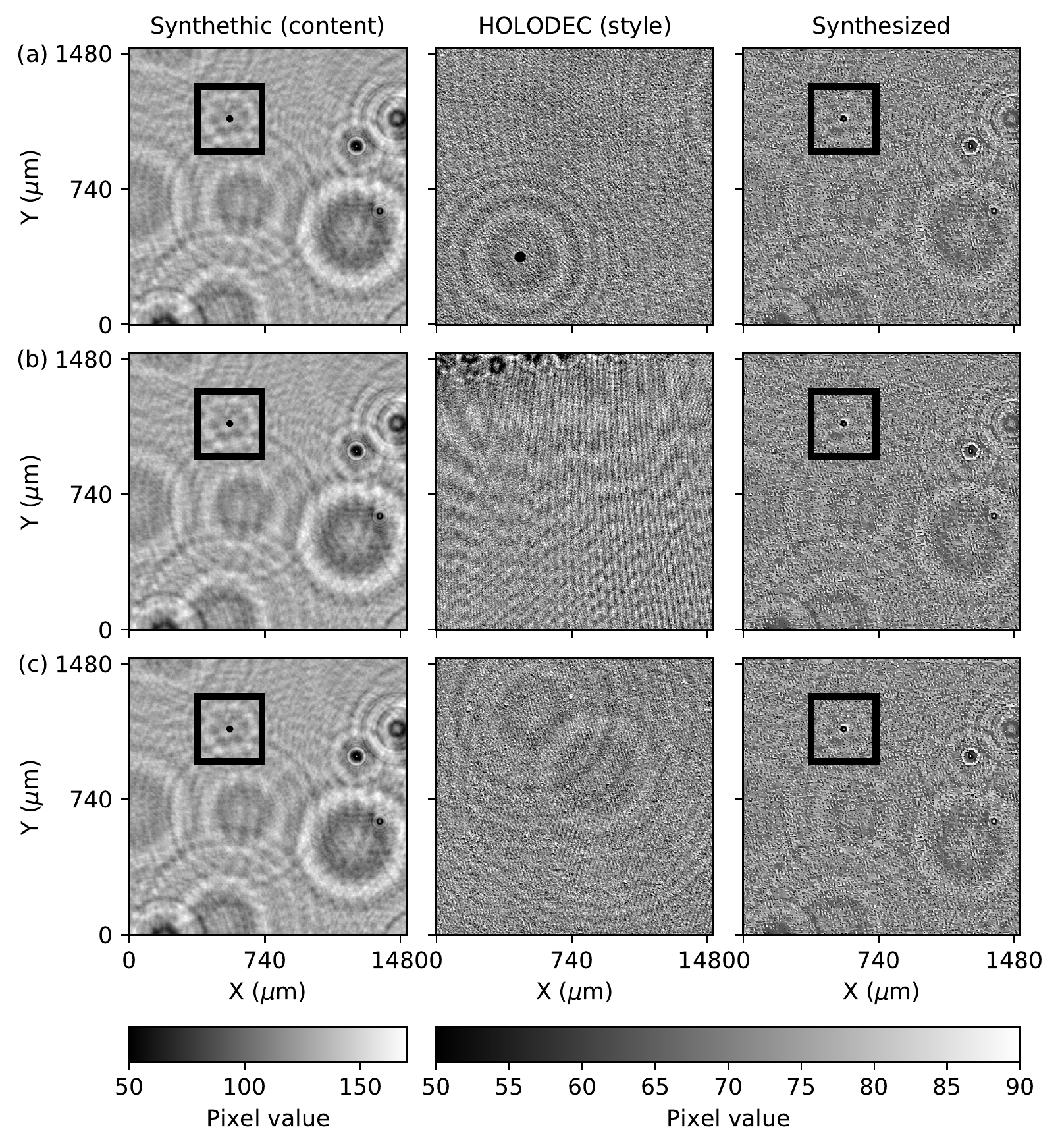}
    \caption{In (a-c), the same synthetic hologram is paired with a different HOLODEC hologram to produce a generated hologram using style transfer. The HOLODEC examples illustrate (a) an in-focus particle, (b) interference patterns and successive dark spots, and (c) small artefacts. U-Nets are observed to perform poorly on (b) and (c) when trained with synthetic images only. The content weight $\alpha$ = 1, while the style weight $\beta$ = $\beta_0$ = $10^9$.}
    \label{fig:hologram_style_example}
\end{figure}

\begin{figure}[t!]
    \centering
    \includegraphics[width=\columnwidth]{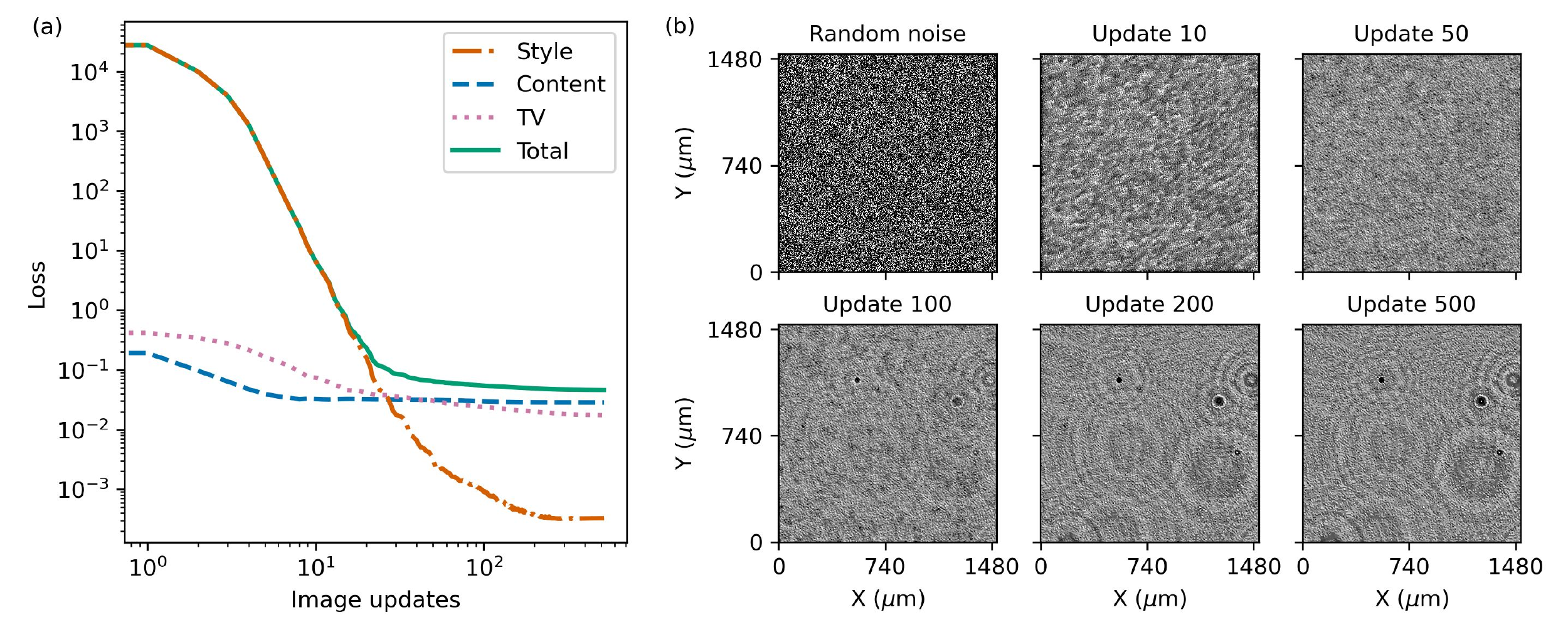}
    \caption{(a) The style, content, TV, and total loss are plotted versus the number of training epochs (the number of times the image was updated). (b) The initial random image and selected images at different training updates are shown. The pixel values shown ranged from 50 to 90, as in Figure~\ref{fig:hologram_style_example} for the HOLODEC and generated images.}
    \label{fig:loss}
\end{figure}

Figure~\ref{fig:hologram_style_example} shows a synthetic hologram (content image) that contains a single in-focus particle, and different HOLODEC hologram examples (style images) that were used to generate the style transferred image, for $\alpha$ = 1, and $\beta$ = $10^9$. The HOLODEC example shown in Figure~\ref{fig:hologram_style_example}(a) was selected because it contains an in-focus particle. Note that while there appears to be a transfer of fine scale texture from the style images, particles from these style images are not transferred generated images.  Thus there is no need to locate empty holograms for the style transfer images or perform careful QC in selecting the appropriate images for this process.

The images in Figure~\ref{fig:hologram_style_example}(b-c) are typical examples where a U-Net trained only on synthetic images (left column) performs poorly on actual HOLODEC images due to a high false-positive rate. The interference fringes seen in the HOLODEC image in Figure~\ref{fig:hologram_style_example}(b) often produce patterns that resemble in-focus particles, while (c) shows an example where many small artefacts are present and have a round shape. In all three examples the in-focus particle in the synthetic example is still present in the generated examples, and the overall generated image contrast is much more comparable to the HOLODEC images. 

Figure~\ref{fig:loss}(a) shows the training curve for the generated image shown in Figure~\ref{fig:hologram_style_example}(a). Figure~\ref{fig:loss}(b) illustrates the progression of the predicted image at selected epochs. Starting initially with a random noise image for $x$, a very steep drop in the style loss is observed over the first few epochs. The content and TV losses also drop quickly after the first few epochs, then the curves mostly flatten after about 10 epochs, and the content contribution comes to dominate the total loss thereafter. In these examples, the choice of $\alpha / \beta$ determined the similarity between the generated image and content and style images used to create it. Here, according to the AlexNet metric, all the generated images are more similar to the HOLODEC images than to the synthetic image. 

\begin{figure}[t!]
    \centering
    \includegraphics[width=\columnwidth]{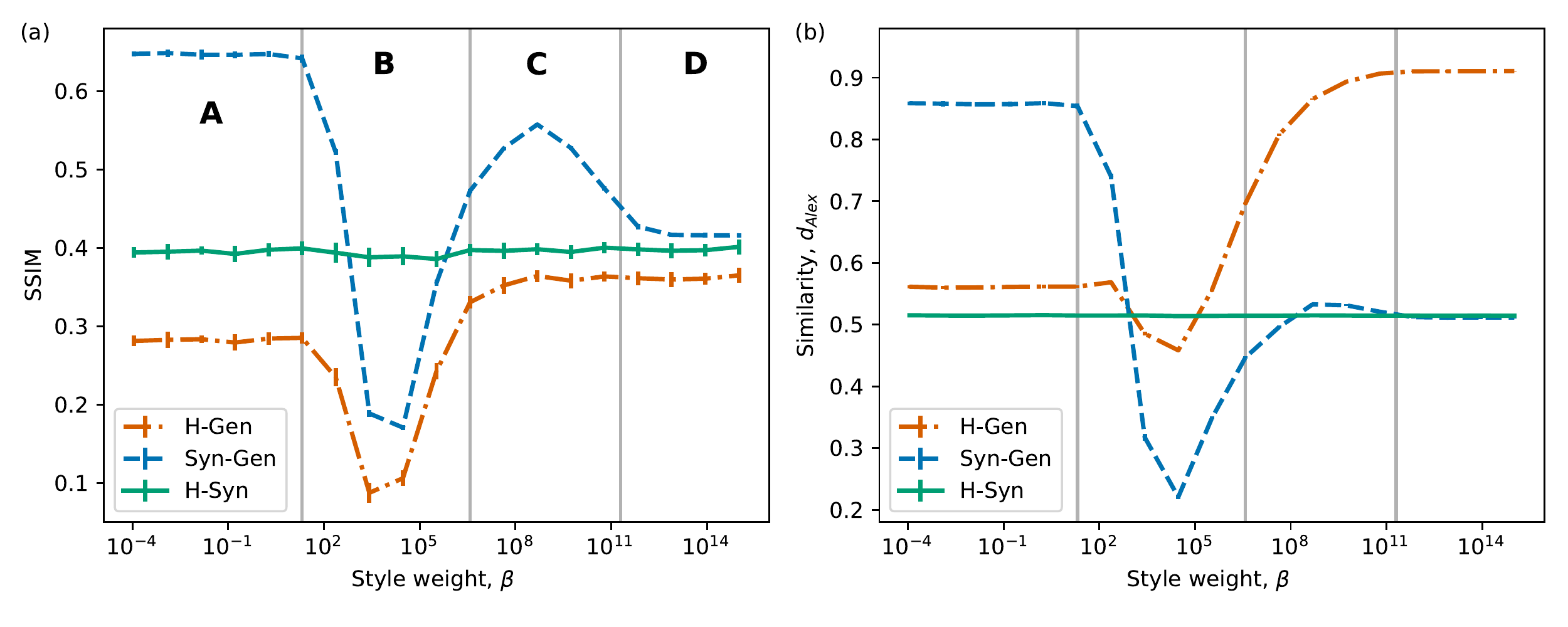}
    \caption{(a) The average SSIM and (b) average $d_\text{Alex}$ scores computed using the test data sets, is shown versus the style weight $\beta$ in Equation~\ref{style_loss} with $\alpha$ and $\gamma$ = 1.} 
    \label{fig:similarity}
\end{figure}

Figure~\ref{fig:similarity}(a) and (b) show the SSIM and $d_\text{Alex}$ metrics versus $\beta$ with $\alpha$ = 1.  These metrics are computed for test data of HOLODEC-synthetic, HOLODEC-generated, and synthetic-generated pairs. Note that these metrics are constant for the HOLODEC-synthetic pairs because these images are unaffected by $\beta$. 

In both figures, four regions of interest are approximately identified with vertical lines. Region A corresponds with values of approximately $\beta \leq 10$ where the content loss dominates the total loss. Accordingly, both metrics show that the style-transferred images strongly resemble the original synthetic images (see also Figure~\ref{fig:hologram_style_example}(a,b)). The main difference between the two metrics is that $d_\text{Alex}$ scores HOLODEC-generated pairs higher relative to HOLODEC-synthetic pairs than does SSIM. However, the curves are all qualitatively similar between the two metrics for all values of $\beta$.

\begin{figure}[t!]
    \centering
    \includegraphics[width=7 in]{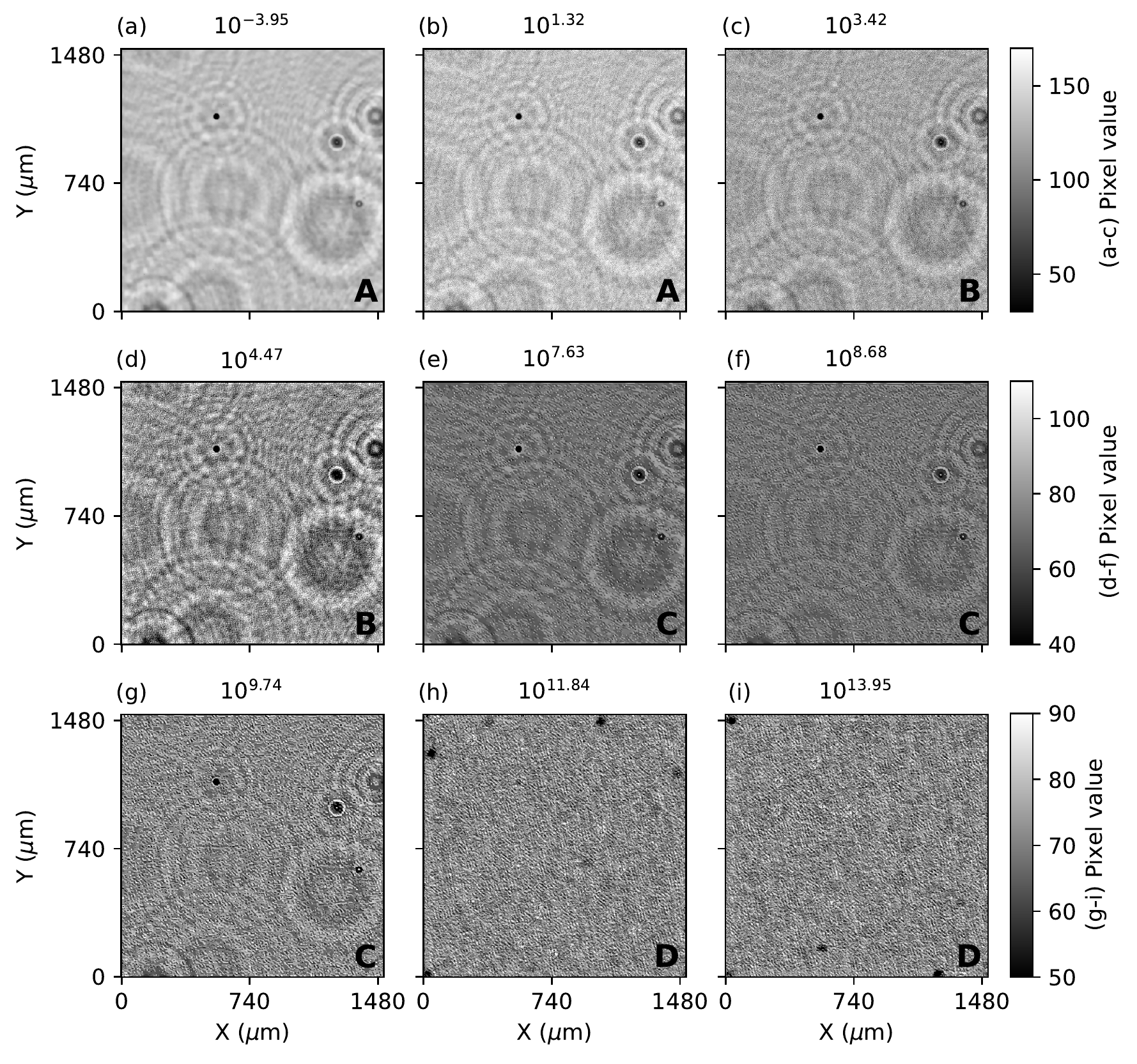}
    \caption{(a-i) Examples of generated holograms using the content and style images from Figure~\ref{fig:hologram_style_example}(a) and different style weights. The region is listed in the bottom right.}
    \label{fig:weight_exmaples}
\end{figure}

Examples of images from each region in Figure~\ref{fig:similarity} are shown in Figure~\ref{fig:weight_exmaples} using the  content and style images from Figure~\ref{fig:hologram_style_example}(a) with $\alpha = 1$ and varying $\beta$.

Region B is identified approximately when $10^2$ < $\beta \le 10^7$, which is when both metrics for synthetic-generated and HOLODEC-generated pairs drop significantly. The examples in Figure~\ref{fig:hologram_style_example}(c,d) show that the generated image starts to look significantly different compared with the original synthetic image as the style term begins to contribute. Although these examples may give the impression that the generated images still look like the synthetic images, they are in reality a mixed representation which does not on average resemble the synthetic or HOLODEC hologram (for example, the average contrast difference for examples like those in Figure~\ref{fig:hologram_style_example}(c,d) is less than that for the synthetic images but still greater than what is observed in the HOLODEC examples). Around $\beta = 10^4$ both metrics reach a minimum before beginning to rise with increasing $\beta$. 

Region C is identified approximately when $10^7 < \beta \le 10^{12}$. Both SSIM and $d_\text{Alex}$ continue to increase for synthetic-generated pairs and peak near $\beta = 10^9$, though the peak is more pronounced for SSIM. Both scores for HOLODEC-generated pairs also increase and reach plateau values, which are higher compared against that in regions A and B, indicating the images have been stylized to some degree while the the Synthetic-Generated scores indicate the image still retains content features. Clearly, the neural $d_\text{Alex}$ determines the generated holograms are more similar to HOLODEC holograms compared to SSIM, but overall both metrics indicate the predicted images have struck a particular balance between retained content and learned style features. Figure~\ref{fig:hologram_style_example}(e-g) show that the in-focus particle remains in place and roughly the same shape, while, for example, the image contrast is adjusted to be more like that of the HOLODEC images.

Finally, region D is identified by $\beta > 10^{12}$, which is when the style term dominates the loss. The $d_\text{Alex}$ score for style-generated pairs remains high while that for content-generated pairs converges to the same value for content-style pairs, which is to say the metric cannot differentiate the generated image from style images. However, Figure~\ref{fig:hologram_style_example}(h,i) show that the in-focus particle and other content from the synthetic image has largely been removed. Furthermore, what appear to be fake particles are now being introduced into the generated images. In other words, the original image was over-stylized. 
 
Region A represents the instances where output has very little style imposed on the content images and therefore most closely resembles idealized synthetic data, which we previously established in \cite{schreck2022} makes a poor U-Net training data set when the objective is to process actual HOLODEC data.  Images from region D represent instances where content is largely absent from output.  Clearly we cannot expect to train a U-Net to recognize particles that are not in the input images.  Without having to resort to manual labeled HOLODEC images, the obvious choice would be to select the value of $\beta$ around $\beta_0=10^9$ in region C, when both metrics for synthetic-generated pairs peak, and the generated examples appear to be more similar to HOLODEC examples according to $d_\text{Alex}$. 

The performance of the U-Net trained on the generated images is then expected to peak at mask reproduction when $\beta$ is around $\beta_0 = 10^9$, as the predicted images are the most similar to the content images in region C. Also, because the generated images are similar to the HOLODEC images in region C relative to A and B, we might expect U-Net models trained on them to perform the best on HOLODEC images. If this is the case, one can select the value of $\beta$ by finding the value where  the generated images are more similar to the HOLODEC images according to $d_{Alex}$ and SSIM.  This selection should result in the best generated dataset for training the U-Net.

\subsection{Performance on synthetic and HOLODEC images}

\begin{figure}[t!]
    \centering
    \includegraphics[width=\columnwidth]{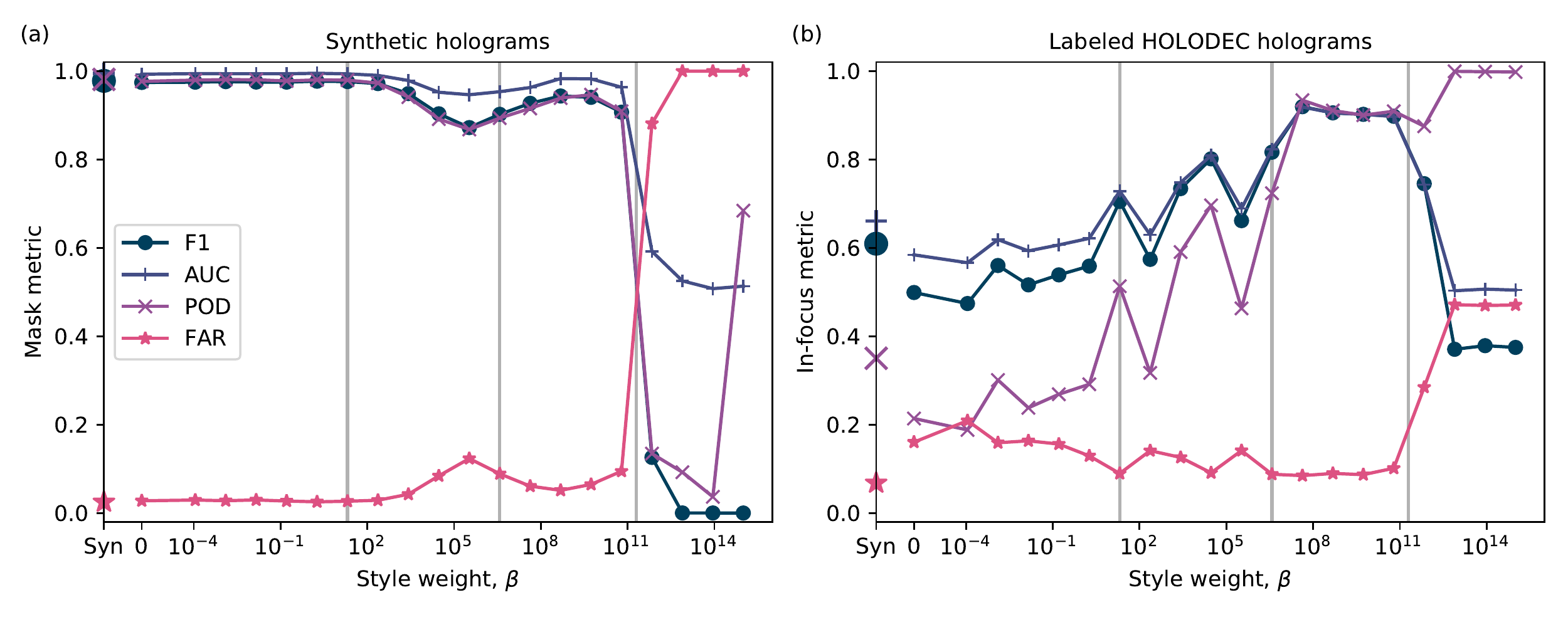}
    \caption{(a) Several binary metrics for the U-Net's mask prediction performance are plotted versus style weight $\beta$ for synthetic or style-generated images. (b) The same binary metrics are shown for the particle detection task in manually labeled  HOLODEC images versus $\beta$. As the models did not rely on the manual examples during training, all 2,356 examples were used in the figure. The x-axis labels "Syn" and "0" refer to U-Nets trained on synthetic images and generated images with $\beta = 0$, respectively.}
    \label{fig:unet_performance}
\end{figure} 

In order to test the hypothesis that the optimal $\beta$ can be found analytically from $d_{Alex}$ and SSIM, the generated images (and their corresponding labels) at the different values of $\beta$ are used to train U-Net models, as is shown schematically in Figure~\ref{fig:unet}(b). The model architecture and training hyperparameters were all identical for each different value of $\beta$ (see Appendix \ref{appendix:model} for more details). Post-training, each U-Net model is run in operation mode on the test set of generated images and the manually labeled test set of actual HOLODEC tiles. Figure~\ref{fig:unet_performance}(a) shows several binary performance metrics for the U-Net at reproducing the mask labels accompanying the generated images, while Figure~\ref{fig:unet_performance}(b) shows similar quantities for particle detection in true HOLODEC tiles. Since none of the manually labeled tile examples were used in the creation of the style-generated images, all 2,356 were used in the figure. The metrics are the the F1 score, the area-under-the-ROC-curve (AUC), the probability of detection (POD), and the false alarm rate (FAR).

Figure~\ref{fig:unet_performance}(a) shows overall gradually diminishing training performance with increased $\beta$.  This is expected as the pristine simulations are increasingly stylized with non-ideal instrument noise and artifacts, making particle detection more difficult.  In region A, when the generated holograms resemble the synthetic examples, the mask prediction performance is very high and generally flat. But then in region B, when the generated holograms are modified more by the style image, there is a clear drop in the AUC and POD while the FAR rises. As $\beta$ continues to grow and the generated images start to resemble both content (synthetic) and style (HOLODEC) examples, the performance rises and reaches a peak around $\beta$ = $\beta_0$ before dropping, first slowly and then very suddenly once in region D. We expect this peak in region C to be less than that in A where the images lack the characteristic noise of the instrument.  In region D, the AUC is about 0.5 which should be expected as the over-stylized images and the mask labels retain very little content from the actual particles.  The trained U-Net outputs all, or too many 1s when trained on over-stylized images, so POD quickly approaches 1. We also computed diameters for the predicted masks and found no significant deviation from the mean true value for the test set, for all choices of $\beta$ except those in region D.

Overall, Figure~\ref{fig:unet_performance}(b) shows that the relevant range of $\beta$ may be selected according to our hypothesis. Relative to the U-Net trained only on synthetic images (labeled `Syn' on the x-axis in Figure~\ref{fig:unet_performance}(b)), the performance on the HOLODEC examples in region A is signified by lower PODs and higher FARs, although the AUC values are all approximately around 0.6.   Even the model trained on generated images with $\beta = 0$ showed a higher FAR relative to 'Syn', indicating that the U-Net trained on idealized synthetic holograms tends to under predict particles, to the benefit of a low FAR. We should note here that the difference in performance between the synthetic trained and $\beta=0$ cases indicates that setting $\beta=0$ does not perfectly reproduce the content image, resulting in slightly lower U-Net performance.  However, neither case is relevant to the objective of this work since both cases fail to perform well on HOLODEC data.  In region B, the particle detection performance generally rises even as the mask performance drops. However, the particle detection performance zig-zags quite significantly making a selection of $\beta$ unreliable without manual labeled examples. In Region C, the detection performance reaches an approximate plateau, with high AUC values and consistently low FAR values. The performance seems to slightly decrease as the style weight approaches values in region D. In region D, the model is unable to train effectively, resulting in an output of all 1s.  This results in a high POD and FAR with AUC at 0.5. Region C models also performed the best overall on the HOLODEC images relative to A, B and D. As expected, models in Region D do not perform at all on the manual examples, as signified by AUCs close to 0.5. 

\subsection{Comparison of optimized models}

\begin{table}[t!]
\begin{center}
\begin{tabular}{ l | c | c | c | c | c | c | c | c | c | c }
\multicolumn{1}{c |}{Metric} & \multicolumn{2}{c |}{F1} & \multicolumn{2}{c |}{AUC} & \multicolumn{2}{c |}{POD} & \multicolumn{2}{c |}{FAR} & \multicolumn{2}{c }{CSI} \\
\hline
 & Syn/Gen & HOLO & Syn/Gen & HOLO & Syn/Gen & HOLO & Syn/Gen & HOLO & Syn/Gen & HOLO \\ 
\hline
Synthetic  & 0.977 & 0.220 & 0.992 & 0.448 & 0.979 & 0.106 & 0.025 & 0.387 & 0.956 & 0.100 \\
Optimized noise & 0.968 & 0.866 & 0.985 & 0.808 & 0.967 & 0.922 & 0.031 & 0.096 & 0.937 & 0.840 \\
Stylized & 0.966 & 0.864 & 0.988 & 0.805 & 0.968 & 0.922 & 0.035 & 0.097 & 0.935 & 0.839 \\ 
\hline\hline
\end{tabular}
\end{center}
\caption{Several metrics are listed for each hyperparameter optimized model and were computed on the synthetic or generated (Syn/Gen), and HOLODEC (HOLO) test data sets, for mask and particle detection (binary) predictions, respectively. The validation set of labeled HOLODEC images was used to optimize the noise parameters in the Optimized noise model.} %
\label{comparison}
\end{table}

Now that the style weight may be selected without relying on manual examples for optimization, we selected $\beta = \beta_0 = 10^9$ and trained a model on the style-transferred images. Table~\ref{comparison} compares this model (`Stylized') versus a model trained on idealized synthetic images only (`Synthetic'), and a model trained on synthetic images with noise introduced during training (`Optimized noise', where the optimization required manual labels). In all three cases we performed extensive hyperparameter optimization to find optimal architectures and training parameters using the mask performance on the validation set of generated images as the optimization metric. The model which leveraged noise during training additionally used the manually labeled validation HOLODEC examples to guide optimization of the added noise. The same binary metrics in Figure~\ref{fig:unet_performance} are listed as well as the critical success index (CSI) for mask performance on test images (either synthetic or generated) and detection performance on test HOLODEC images. As the validation set of manually labeled examples were used to select the noise added which resulted in the Optimized noise, only the testing split is used in Table~\ref{comparison}.

Table~\ref{comparison} shows that the Stylized model and the Optimized noise model gave nearly the same performance on all metrics on both data sets even though the Stylized approach does not require manual labeling. The Optimized noise model outperforms Stylized by hundredths of a percent in each metric category. The Synthetic model showed better mask performance on synthetic data. However, the Synthetic demonstrated much worse performance at particle detection on real HOLODEC data compared to the other two models. For example Table~\ref{comparison} shows that it had a much higher FAR, and the CSI was more than 70\% lower compared to the other models.

\begin{figure}[t!]
    \centering
    \includegraphics[width=\columnwidth]{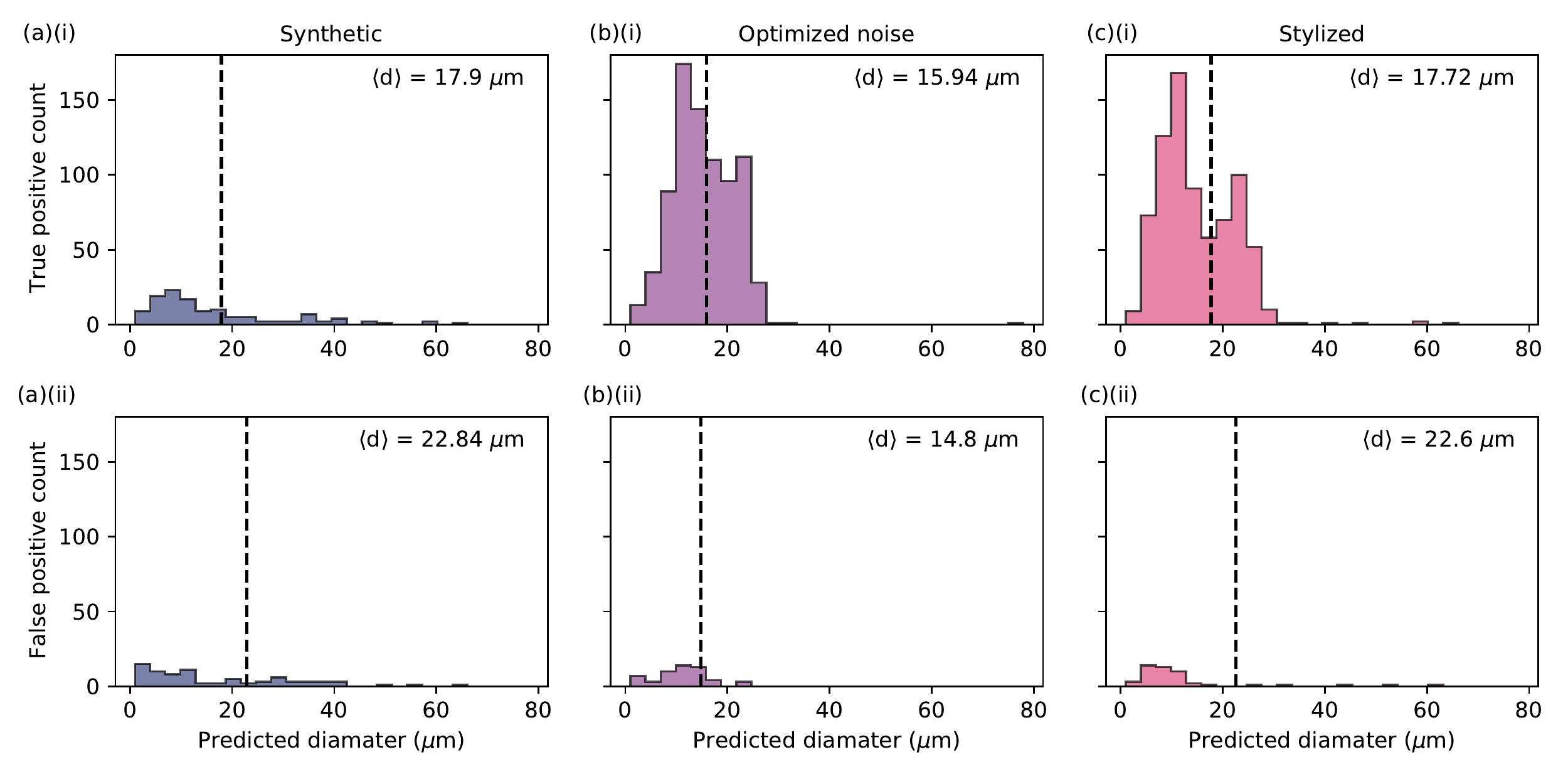}
    \caption{(a-c) Histograms of computed particle diameters for the three models for the test set of manually labeled HOLODEC images. The dashed vertical line in each panel denotes the predicted average particular diameter $\langle d \rangle$.}
    \label{fig:diameters}
\end{figure}

Lastly, as the Optimized noise and Stylized models produced nearly identical performance, the diameters for the predicted particles in the test set of manually labeled HOLODEC examples are computed to probe how the models describe the particles shapes. Figure~\ref{fig:diameters} shows histograms of the predicted diameters for the three models. A particle diameter was computed for a predicted mask if at least one in-focus particle was present. From the predicted mask, any pixel labeled 1 was grouped with a neighboring pixel along $x$ or $y$ (but not along the diagonal in the plane) if it was also labeled 1. Therefore, breaks between groups defines multiple particles. The diameter of a group was taken to be it's maximum extent in either x or y.

Figure~\ref{fig:diameters}(a)(i) shows the Synthetic model expectantly under-predicting the true particles compared with the other two models, and nearly as many false positive predictions were made relative to true positive, as seen in Figure~\ref{fig:diameters}(a)(ii). Meanwhile, the Optimized noise and Stylized models predicted similar distributions, but which differed from the Synthetic distribution.  Both models predicted two peaks centered approximately around 15 and 22 \textmu m, respectively, with the Stylized model having the more pronounced second peak. The Stylized model also predicts slightly greater numbers of larger particles relative to Optimized noise that are also sometimes observed by the Synthetic model. As a result, the overall average diameter is predicted to be about 17.8 \textmu m versus 15.9 \textmu m for Stylized and Optimized noise models, respectively. Figure~\ref{fig:diameters}(b)(ii) and (c)(ii) further show that both models mostly predict false diameters at 15 \textmu m or smaller. 

\section{Discussion}
\label{discussion}

Overall, the performances observed on both synthetic and HOLODEC images with the Optimized noise and the Style model were comparable, with both models performing well against models trained on the synthetic holograms only. This shows the advantage of using the style-transfer method for hologram image augmentation, which obviated the need to perform manual labeling as a means for optimizing noise added during the training of U-Net models. Additionally, we did not have to ambiguously choose the types of augmentations performed on the images, as was required for the Optimized model. Furthermore, the application of style transfer to images that were used for training, rather than applied during operation, means that computational performance is also comparable with the Optimized noise model. 

The main computational bottleneck involving the style-transfer method occurs during the creation of style-transferred training data sets. Many generated images needed to be created, one at a time, from a unique content-style image pair, that all required the iterative optimization described above. The image sizes used were also quite large. That further slowed training times and made it cumbersome to create large data sets used for exploring the different weight combinations. Improved approaches could be focused on replacing the optimization with a trained neural network that predicts the stylized image. This can lead to several orders of magnitude speed-up over the original method when feed-forward architectures are utilized \citep{li2016, ulyanov2016, huang2017}. These may include the StyleGAN method \citep{karras2021}, which integrates the concepts of style transfer with a generative adversarial network \citep{goodfellow} and allows for fine-tune control over the relative strength of image features at different scales. For holograms, the StyleGAN could potentially be used to more selectively control specific content or style features on demand (for example, adding more features pertaining to artefacts).

We also only worked with CSET holograms, in particular the RF07 data split, which had limited particle densities as well as sizes. How well a trained model, which leveraged style images from one field campaign, works on other campaigns remains to be tested in future work. Style-generated hologram data sets, which do not resemble operational inputs, plausibly would not be feasible as training targets, as we learned in \cite{schreck2022}. In such a scenario, the holograms from the latest field campaign could be used as style examples to generate the relevant training data set. The StyleGAN approach again could potentially take advantage of all the different types of holograms obtained from field campaigns (and all of the unique objects contained in each one), since it may be capable of leveraging multiple styles and content features arising in the data sets.

Finally, as the primary objective of this study was to remove the manual-labeling step without sacrificing performance, we did not try to assess the physical realness of the generated holograms since during operation only real HOLODEC holograms are used. However, the question remains as to whether they are physically reasonable. One way to test this assumption could be to perform style-transfer using full-size holograms, which would be computationally challenging but not impossible with current GPUs, and then refocus them via wave-propagation along the focal plane. This would require faithfully reconstructing the real and imaginary components of the electric field, rather than just the intensity, as was done here since style-transfer was applied to tiles selected from the full-sized holograms after they had been propagated.

\section{Conclusions}
\label{conclusions}

In summary, we have shown that the style-transfer algorithm is an effective approach for translating synthetic holograms, which were created using an idealized physical model of the instrument, into holograms that resemble those actually observed by the instrument. In principle, the application of style-transfer to perform the image augmentations should not be limited to holograms since non-ideal instrument behavior is a problem across many domains. When the synthetic holograms are used in machine learning models to predict masks around in-focus particles as was done here, noise had to be injected onto the images during training so that the model performed well on both synthetic and raw data. However, the choice of noise transformations and parameters can only be selected after expensive manual labeling of raw images and hyperparameter optimization. The style-transfer approach, which transformed the clean hologram data set into one more resembling observed holograms, delivered the same mask prediction performance without the cumbersome requirement of manual labeling. Furthermore, both models had comparable performance on a small set of manually labeled HOLODEC examples, and predicted similar distributions for the particle diameters.

\appendix

\appendixfigures  
\appendixtables   

\subsection{Model and training hyperparameters}
\label{appendix:model}

\begin{table}
\begin{center}
\begin{tabular}{ c | c | c | c | c }
\hline
Parameter & Synthetic &  Optimized noise & Stylized ($\beta_0$) & $10^{-5}$ < $\beta$ $\le$ $10^{15}$\\ 
\hline
Learning rate      & \SI{3.86e-4}{} & \SI{2.46e-4}{} & \SI{8.5e-5}{} & \SI{1e-3}{} \\
Training loss      & Focal-Tyversky &  Focal-Tyversky & IoU & Focal-Tyversky \\
Segmentation model & U-Net &  LinkNet & U-Net & U-Net \\
Encoder model      & EfficientNet-b0 &  Xception & DenseNet121 & VGG11 \\
Tile transform     & 255 &  Normalized & Standard & Symmetric \\ 
L2 regularization  & \SI{0.0}{} & \SI{0.0}{} & \SI{2.1e-7}{} & \SI{6.0e-6}{} \\
Gaussian blur $\sigma$ & - & 2.125 & - & - \\
Gaussian noise         & - & 0.326 & - & - \\
Brightness factor      & - & 1.270 & - & - \\ 
\hline\hline
\end{tabular}
\end{center}
\caption{The values of the best hyperparameters in the optimization studies for the neural segmentation models for the three species. The batch size was fixed at 16.}
\label{best_parameters}
\end{table}
 
 Table~\ref{best_parameters} lists the best parameters found for the segmentation models investigated. See reference \cite{schreck2022} for more details on the hyperparameter optimization of segmentation models with hologram data sets. The segmentation models found to be optimal in Table~\ref{best_parameters} were the U-Net \citep{ronneberger2015u} and LinkNet \citep{chaurasia2017linknet}, while the pre-trained encoder model weights considered were DenseNet-121 \citep{huang2017densely}, Xception \citep{chollet2017xception}, EfficientNet-b0 \citep{tan2019efficientnet}, and VGG-11 \citep{simonyan2014very}. See the package segmentation-models-pytorch located at \url{https://github.com/qubvel/segmentation_models.pytorch} for more details on the segmentation and encoder models. The optimal training losses found were the intersection over union (IOU) and Focal-Tyversky losses \citep{lin2017focal,salehi2017tversky}. For additional definitions of each loss function, see the Holodec-ML software package located at \url{https://github.com/NCAR/holodec-ml}. We also utilized pre-trained weights obtained from the ImageNet data set in all trained models. The tile transforms were applied to a tile just before being passed into a segmentation model include dividing all pixels by 255, re-scaling all pixels to lie between 0 and 1 (Normalized) or -1 and 1 (Symmetric), or re-scaling each tile by subtracting the pixel mean and dividing by the square of the variance (Standard). 
 
 \subsection{Structural similarity index metric (SSIM)}
 \label{app:ssim}
 
 The global structure similarity definition for comparing images $x$ and $y$ is given by
 \begin{equation}
 \label{eq:ssim}
     SSIM(x, y) = \frac{(2 \mu_x \mu_y + C_1)(2 \sigma_{xy} + C_2)}{(\mu_x^2 + \mu_y^2 + C_1)(\sigma_x^2 + \sigma_y^2 + C_2)} \\ 
 \end{equation}
 where $C_1$ and $C_2$ are fixed constants, and $\mu$ and $\sigma$ are defined as 
\begin{equation}
    \mu_x = \frac{1}{N} \sum_{i=1}^N x_i \\ 
    \sigma_x^2 = \frac{1}{N-1} \sum_{i=1}^N (x_i - \mu_x)^2 \\ 
    \sigma_{xy} = \frac{1}{N-1} \sum_{i=1}^N (x_i - \mu_x) (y_i - \mu_y).
\end{equation}
In practice, we use the mean structural similarity index which averages over an 11 by 11 circular-symmetric Gaussian Weighting function \citep{ssim}.
 
\subsection{Data sets}
The HOLODEC and synthetic data sets can be accessed at \url{https://doi.org/10.5281/zenodo.6347222}. The labeled HOLODEC examples were a subset of the RF07 data set (validation set ID 0-9, testing set ID 10-19), while the synthetic holograms were generated with simulations. All holograms in the RF07 data set were used for creating the style-transferred data sets. See Section~\ref{datasets} for more details. The style generated data sets can be built using the Holodec-ML software package located at \url{https://github.com/NCAR/holodec-ml}.

\noappendix  

\begin{acknowledgements}
This material is based upon work supported by the National Center for Atmospheric Research, which is a major facility sponsored by the National Science Foundation under Cooperative Agreement No. 1852977.  We would like to acknowledge high-performance computing support from Cheyenne and Casper \cite{Cheyenne} provided by NCAR's Computational and Information Systems Laboratory, sponsored by the National Science Foundation. The neural networks described here and simulation code used to train and test the models are archived at \url{https://github.com/NCAR/holodec-ml}. All HOLODEC and synthetic hologram data sets created for this study are available at \url{https://doi.org/10.5281/zenodo.6347222}
\end{acknowledgements}

\bibliographystyle{copernicus}
\bibliography{references}

\end{document}